\documentclass{iau379}

\usepackage{graphicx}
\usepackage{amsmath}
\usepackage{multirow}

\begin{document}

\lefttitle{M\"uller}
\righttitle{IAU Symposium 379: Template}

\jnlPage{1}{7}
\jnlDoiYr{2023}
\doival{10.1017/xxxxx}

\aopheadtitle{Proceedings of IAU Symposium 379}
\editors{P. Bonifacio,  M.-R. Cioni, F. Hammer, M. Pawlowski, and S. Taibi, eds.}

\title{Planes of satellites in the nearby Universe}

\author{M\"uller O. $^1$}
\affiliation{$^1$Institute of Physics, Laboratory of Astrophysics, Ecole Polytechnique Fédérale de Lausanne (EPFL), 1290 Sauverny, Switzerland}

\begin{abstract}
Since the mid 70ies it is known that the dwarf galaxies around the Milky Way are arranged in a thin, polar structure. The arrangement and motion within this structure has been identified as a severe challenge to the standard model of cosmology, dubbed as the plane of satellites problem. More observational evidence for such structures has been put forward around other galaxies, such as the Andromeda galaxy, Cen\,A or NGC\,253, among others, adding to the previously identified tensions. Solutions to the plane of satellite problem should therefore  not only be tailored to the Milky Way, but need to explain all these different observed systems and environments. 
\end{abstract}


\maketitle

\section{Introduction}
The plane of satellites problem \citep{2018MPLA...3330004P} is one of the most outstanding and heavily debated problems in near-field cosmology. It describes a discrepancy between the observed and predicted arrangement and motion of dwarf galaxy systems. In observations, dwarf galaxy systems around their host galaxies seem to be flattened and co-moving, while a more isotropic distribution and random motions is found in cosmological simulations. Together, these two feature -- distribution and motion -- consists of the plane of satellite problem. It has been argued that this configuration of a satellite system is in severe tension with the standard model of cosmology based on statistical arguments, finding less than 1\% of analogs in cosmological systems hosting such arrangements \citep{2014ApJ...784L...6I}. Others argued that the frequency in cosmological simulations is much larger, at the 10\% level \citep{2015MNRAS.452.3838C}. It all depends on what is to be considered an analog of an observed system -- no consensus has been established to date with different research teams using different definitions of what an analog of an observed satellite system is.

While there is still a disagreement of the actual tension within the cosmological paradigm, progress has been made on the observational side, with the plane of satellites problem not being constrained to the Local Group anymore. For example was it noted by \citet{2013AJ....146..126C} that the close-by M\,81 satellite system exhibits some flattening, orientated within the Local Sheet. \citet{2015ApJ...802L..25T} found that around Cen\,A, another neighbor of the Local Group, the dwarfs seem to be arranged in two parallel planar structures. Therefore, the plane of satellite problem  must be considered in a wider context.

\section{Observed planar structures}
In the following I will review the observational evidence for planes of satellites in the nearby universe. 

\subsection{The Local Group planes}
The best evidence for planes of satellites we have in the Local Group, namely for the Milky Way and the Andromeda galaxy. For the Milky Way, \citet{1976MNRAS.174..695L} found that the then known dwarf galaxies (LMC, SMC, Draco, Ursa Minor, Sculptor), as well as the globular cluster Palomar 13 and the Magellanic stream, are arranged in a thin, polar structure. Independently, \citet{1976RGOB..182..241K} found a slightly different plane (with an angle of more than 30 degrees from that of \citealt{1976MNRAS.174..695L}), including Palomar 3 and 4 and the distant dwarf galaxy Leo-I. Their plane, however, didn't include Palomar 13, nor the Magellanic stream.  \citet{1976MNRAS.174..695L,1983IAUS..100...89L} argued that the plane might of tidal origin and therefore should include the Magellanic stream. The discovery of dwarf galaxies around the Milky Way \citep[e.g., ][]{2005ApJ...626L..85W,2007ApJ...654..897B,2007ApJ...662L..83W} strengthened the picture of the existence of \citet{1976MNRAS.174..695L}'s planar structure \citep{2005A&A...431..517K,2007MNRAS.374.1125M,2009MNRAS.394.2223M}. Thanks to proper motion measurements of dwarf galaxies \citep[e.g., ][]{2006AJ....131.1445P,2007AJ....133..818P}, it became evident that the dwarfs are not only arranged in a thin plane, but that they also co-orbit within this plane \citep{2008ApJ...680..287M,2012MNRAS.423.1109P,2020MNRAS.491.3042P}. This structure is now dubbed the Vast Polar Structure (VPOS).

For the Andromeda galaxy,  \citet{2006AJ....131.1405K} found a planar alignment for a subsample of the dwarf galaxies, namely for the dwarf ellipticals. They note this arrangement seems alignment with the large-scale structure and polar to the disk of M\,31. At the same time \citet{2006MNRAS.365..902M} found a planar alignment of a subpopulation of dwarf galaxies with respect to the disk of M\,31. New data from the PANDAS survey \citep{2011ApJ...732...76R,2013ApJ...776...80M} shed some light on the distribution of the system. Indeed, there seems to be a co-moving plane of satellites around M\,31, which is neither aligned nor polar with its disk. The co-motion was established using line-of-sight velocities of 15 satellites, of which 13 seem to be consistent with a co-rotational signal \citep{2013Natur.493...62I}. \citet{2020ApJ...901...43S} measured the proper motion of two satellites within this plane and found that they are aligned with it, which is further evidence for the co-rotating nature of the structure, however, observations of the remaining dwarfs are needed to further assess this situation. \citet{2020MNRAS.499.3755S} found another planar alignment of dwarf galaxies around M\,31 using a 4-galaxy-normal density plot \citep{2013MNRAS.435.1928P}. This plane is seen face-on from our point of view and consists of 39 dwarf galaxies. Based on its velocity dispersion, it should not be stable and disperse eventually.

\subsection{The Centaurus Group}
The Centaurus Group is a nearby galaxy aggregate consisting of two giant galaxies -- the beautiful Cen\,A galaxy at 3.8\,Mpc \citep{2010PASA...27..457H} and the grand spiral M\,83 at 4.8\,Mpc \citep{2007AJ....133..504K}. \citet{2015ApJ...802L..25T} argued based on the three dimensional positions of the dwarfs around Cen\,A that they are arranged in two thin planes. Discoveries of more dwarfs around Cen\,A \citep{2014ApJ...795L..35C,2016ApJ...823...19C,2017A&A...597A...7M} made this distinction between the two thin planes not as clear anymore \citep{2016A&A...595A.119M}. \citet{2018Sci...359..534M} studied the kinematics of the dwarf galaxy system and found that 14 out of 16 dwarf galaxies with available line-of-sight velocities were following a co-motion, consistent with a co-rotation within a plane. These numbers have been updated with recent MUSE observations  \citep{2021A&A...645A..92M} to 21 out of 28 dwarf galaxies co-moving \citep{2021A&A...645L...5M}. Comparing the Cen\,A satellite system to cosmological simulations finds similar disagreement between observations and predictions  as for the Milky Way \citep{2021A&A...645L...5M}. What is intriguing to note is that the flattened structure around Cen\,A -- be it one thick or two thin planes -- is aligned with the cosmic web \citep{2019MNRAS.490.3786L}, with its orbital pole pointing towards the Local Void (i.e. the plane is aligned with the Local Sheet).

Around M\,83, \citet{2018A&A...615A..96M} found evidence for another flattening of the satellite system. However, this was based on a smaller number of satellites, with larger uncertainties than that for Cen\,A. Because the distance uncertainties are as large as the putative planar structure, it is not clear whether it is physical or an artefact from the measurement errors.

\subsection{The Sculptor Group}
The Sculptor Group is another galaxy aggregate in our local neighborhood at 3.9\,Mpc \citep{2003A&A...404...93K}. Compared to the Centaurus Group, it is however a very spares environment. The dominant galaxy is the spiral galaxy NGC\,253. Again, different surveys have targeted the vicinity of NGC\,253 to search for dwarf galaxies and follow them up \citep{2014ApJ...793L...7S,2020A&A...644A..91M,2021A&A...652A..48M,2022ApJ...926...77M}. \citet{2021A&A...652A..48M} found a thin spatial alignment of nine dwarf galaxies based on their two and three dimensional positions. Only a handful (i.e. five) of these galaxies have velocities, but a co-moving trend is apparent, with four out of five being consistent with a co-rotational signal. Follow-up observations of their velocities are needed to put this on a more robust data basis.

\subsection{The NGC2750 group}
The NGC\,2750 group is at a larger distance ($\approx$40\,Mpc) than the previously discussed groups, so it represents quite the different environment than that of our own Local Group and its neigbors. \citet{2021ApJ...917L..18P} found evidence for a co-moving dwarf galaxy system, based on the motion of six satellites. Compared to the other systems presented before, the NGC\,2750 dwarf galaxy system is not a flattened structure and more reminiscent of the findings by \citet{2014Natur.511..563I} on the anti-correlation of diametrically opposed galaxy satellites. It certainly will be interesting to follow this group up and search for more dwarf galaxies, as well as get more velocities.

\subsection{The MATLAS survey}
A more statistical approach was done in the MATLAS survey \citep{2020MNRAS.491.1901H,2021MNRAS.506.5494P,2021A&A...654A.105M}. This MegaCam based survey targeted over 200 early-type galaxies between 10 to 45\,Mpc, around which over 2000 dwarf galaxies were discovered \citep{2020MNRAS.491.1901H}. \citet{2021A&A...654A.161H} used this catalog of dwarf galaxies to systematically measure the flattening of the satellite systems. Because only two dimensional information is available, the analysis was performed using the on-sky distribution of the dwarf galaxies. \citet{2021A&A...654A.161H} found that around 30\% of their satellite systems exhibit a significant flattening, more than what is expected from random distributions. However, none of the systems has enough line-of-sight velocities to study the kinematics of the dwarf galaxies, so it is not clear if there is co-motion of the dwarf galaxies as found for the e.g. Cen\,A. Following up the MATLAS dwarfs will be a major undertaking, which is currently underway (\citealt{2021ApJ...923....9M}, Heesters et al., in preparation).

\section{A challenge or not?}
The question about whether the plane of satellites problem is a challenge for $\Lambda$CDM cosmology was recently discussed in several papers. \citet{2021NatAs...5.1185P} reviewed the situation for the three best studied systems -- the Milky Way, tha Andromeda galaxy, and Cen\,A -- and concluded that there is currently no satisfying solution withing standard cosmology to explain all the observations. \citet{2022NatAs...6..897S} classified the plane of satellites problem as a severe tension for cosmology. Shortly after this publication, \citet{2023NatAs.tmp...59S} commented that the plane of satellites problem is no longer in tension with $\Lambda$CDM based on the work of \citet{2022NatAs.tmp..273S}, which was, however, focusing only on the Milky Way satellite system.  \citet{2023arXiv230300441X} argued that for the right environment, the Milky Way satellite system may be explained, but also cautioned that to fully solve the plane of satellites problem the other observed planes of satellites also need to be taken into account. 

Because the discussion is mainly focused on the Milky Way and its satellite system -- which is indeed a well-studied system, but still consists of a sample size of one when comparing to cosmological simulations -- it is imperative to study other systems as well. Fortunately, we live in an era of wide field surveys which will unlock the study of dwarf galaxy systems around hundreds of Milky Way analogs. However, we also need to be prepared for this incoming data and have the right tools to compare the observations to cosmology to make meaningful assessments of the status of this lively debated topic.

\bibliographystyle{iaulike}


\begin{thebibliography}{}

\bibitem[Belokurov et al.(2007)]{2007ApJ...654..897B} Belokurov, V., Zucker, D.~B., Evans, N.~W., et al.\ 2007, APJ, 654, 897. 


\bibitem[Cautun et al.(2015)]{2015MNRAS.452.3838C} Cautun, M., Bose, S., Frenk, C.~S., et al.\ 2015, MNRAS, 452, 3838.

\bibitem[Chiboucas et al.(2013)]{2013AJ....146..126C} Chiboucas, K., Jacobs, B.~A., Tully, R.~B., et al.\ 2013, AJ, 146, 126. 

\bibitem[Crnojevi{\'c} et al.(2014)]{2014ApJ...795L..35C} Crnojevi{\'c}, D., Sand, D.~J., Caldwell, N., et al.\ 2014, APJL, 795, L35. 

\bibitem[Crnojevi{\'c} et al.(2016)]{2016ApJ...823...19C} Crnojevi{\'c}, D., Sand, D.~J., Spekkens, K., et al.\ 2016, APJ, 823, 19. 

\bibitem[Habas et al.(2020)]{2020MNRAS.491.1901H} Habas, R., Marleau, F.~R., Duc, P.-A., et al.\ 2020, MNRAS, 491, 1901. 


\bibitem[Harris et al.(2010)]{2010PASA...27..457H} Harris, G.~L.~H., Rejkuba, M., \& Harris, W.~E.\ 2010, PASA, 27, 457. 

\bibitem[Heesters et al.(2021)]{2021A&A...654A.161H} Heesters, N., Habas, R., Marleau, F.~R., et al.\ 2021, A\&A, 654, A161. 


\bibitem[Ibata et al.(2013)]{2013Natur.493...62I} Ibata, R.~A., Lewis, G.~F., Conn, A.~R., et al.\ 2013, Nature, 493, 62. 

\bibitem[Ibata et al.(2014a)]{2014ApJ...784L...6I} Ibata, R.~A., Ibata, N.~G., Lewis, G.~F., et al.\ 2014, APJL, 784, L6. 

\bibitem[Ibata et al.(2014b)]{2014Natur.511..563I} Ibata, N.~G., Ibata, R.~A., Famaey, B., et al.\ 2014, Nature, 511, 563.


\bibitem[Karachentsev et al.(2003)]{2003A&A...404...93K} Karachentsev, I.~D., Grebel, E.~K., Sharina, M.~E., et al.\ 2003, A\&A, 404, 93. 

\bibitem[Karachentsev et al.(2007)]{2007AJ....133..504K} Karachentsev, I.~D., Tully, R.~B., Dolphin, A., et al.\ 2007, AJ, 133, 504. 


\bibitem[Koch \& Grebel(2006)]{2006AJ....131.1405K} Koch, A. \& Grebel, E.~K.\ 2006, AJ, 131, 1405. 


\bibitem[Kroupa et al.(2005)]{2005A&A...431..517K} Kroupa, P., Theis, C., \& Boily, C.~M.\ 2005, A\&A, 431, 517. 


\bibitem[Kunkel \& Demers(1976)]{1976RGOB..182..241K} Kunkel, W.~E. \& Demers, S.\ 1976, The Galaxy and the Local Group, 182, 241

\bibitem[Libeskind et al.(2019)]{2019MNRAS.490.3786L} Libeskind, N.~I., Carlesi, E., M{\"u}ller, O., et al.\ 2019, MNRAS, 490, 3786. 


\bibitem[Lynden-Bell(1976)]{1976MNRAS.174..695L} Lynden-Bell, D.\ 1976, MNRAS, 174, 695. 

\bibitem[Lynden-Bell(1983)]{1983IAUS..100...89L} Lynden-Bell, D.\ 1983, Internal Kinematics and Dynamics of Galaxies, 100, 89

\bibitem[Marleau et al.(2021)]{2021A&A...654A.105M} Marleau, F.~R., Habas, R., Poulain, M., et al.\ 2021, A\&A, 654, A105. 


\bibitem[Martin et al.(2013)]{2013ApJ...776...80M} Martin, N.~F., Ibata, R.~A., McConnachie, A.~W., et al.\ 2013, APJ, 776, 80. 

\bibitem[Mart{\'\i}nez-Delgado et al.(2021)]{2021A&A...652A..48M} Mart{\'\i}nez-Delgado, D., Makarov, D., Javanmardi, B., et al.\ 2021, A\&A, 652, A48. 


\bibitem[McConnachie \& Irwin(2006)]{2006MNRAS.365..902M} McConnachie, A.~W. \& Irwin, M.~J.\ 2006, MNRAS, 365, 902. 


\bibitem[Metz et al.(2007)]{2007MNRAS.374.1125M} Metz, M., Kroupa, P., \& Jerjen, H.\ 2007, MNRAS, 374, 1125. 

\bibitem[Metz et al.(2008)]{2008ApJ...680..287M} Metz, M., Kroupa, P., \& Libeskind, N.~I.\ 2008, APJ, 680, 287. 

\bibitem[Metz et al.(2009)]{2009MNRAS.394.2223M} Metz, M., Kroupa, P., \& Jerjen, H.\ 2009, MNRAS, 394, 2223. 

\bibitem[Mutlu-Pakdil et al.(2022)]{2022ApJ...926...77M} Mutlu-Pakdil, B., Sand, D.~J., Crnojevi{\'c}, D., et al.\ 2022, APJ, 926, 77. 


\bibitem[M{\"u}ller et al.(2016)]{2016A&A...595A.119M} M{\"u}ller, O., Jerjen, H., Pawlowski, M.~S., et al.\ 2016, A\&A, 595, A119. 


\bibitem[M{\"u}ller et al.(2017)]{2017A&A...597A...7M} M{\"u}ller, O., Jerjen, H., \& Binggeli, B.\ 2017, A\&A, 597, A7. 

\bibitem[M{\"u}ller et al.(2018a)]{2018Sci...359..534M} M{\"u}ller, O., Pawlowski, M.~S., Jerjen, H., et al.\ 2018, Science, 359, 534. 

\bibitem[M{\"u}ller et al.(2018b)]{2018A&A...615A..96M} M{\"u}ller, O., Rejkuba, M., \& Jerjen, H.\ 2018, A\&A, 615, A96. 

\bibitem[M{\"u}ller \& Jerjen(2020)]{2020A&A...644A..91M} M{\"u}ller, O. \& Jerjen, H.\ 2020, A\&A, 644, A91. 


\bibitem[M{\"u}ller et al.(2021a)]{2021A&A...645A..92M} M{\"u}ller, O., Fahrion, K., Rejkuba, M., et al.\ 2021, A\&A, 645, A92. 

\bibitem[M{\"u}ller et al.(2021b)]{2021A&A...645L...5M} M{\"u}ller, O., Pawlowski, M.~S., Lelli, F., et al.\ 2021, A\&A, 645, L5. 

\bibitem[M{\"u}ller et al.(2021c)]{2021ApJ...923....9M} M{\"u}ller, O., Durrell, P.~R., Marleau, F.~R., et al.\ 2021, APJ, 923, 9.


\bibitem[Paudel et al.(2021)]{2021ApJ...917L..18P} Paudel, S., Yoon, S.-J., \& Smith, R.\ 2021, APJL, 917, L18. 


\bibitem[Pawlowski et al.(2012)]{2012MNRAS.423.1109P} Pawlowski, M.~S., Pflamm-Altenburg, J., \& Kroupa, P.\ 2012, MNRAS, 423, 1109. 

\bibitem[Pawlowski et al.(2013)]{2013MNRAS.435.1928P} Pawlowski, M.~S., Kroupa, P., \& Jerjen, H.\ 2013, MNRAS, 435, 1928. 

\bibitem[Pawlowski(2018)]{2018MPLA...3330004P} Pawlowski, M.~S.\ 2018, Modern Physics Letters A, 33, 1830004. doi:10.1142/S0217732318300045

\bibitem[Pawlowski \& Kroupa(2020)]{2020MNRAS.491.3042P} Pawlowski, M.~S. \& Kroupa, P.\ 2020, MNRAS, 491, 3042. 

\bibitem[Pawlowski(2021)]{2021NatAs...5.1185P} Pawlowski, M.~S.\ 2021, Nature Astronomy, 5, 1185. 




\bibitem[Piatek et al.(2006)]{2006AJ....131.1445P} Piatek, S., Pryor, C., Bristow, P., et al.\ 2006, AJ, 131, 1445. 

\bibitem[Piatek et al.(2007)]{2007AJ....133..818P} Piatek, S., Pryor, C., Bristow, P., et al.\ 2007, AJ, 133, 818. 

\bibitem[Poulain et al.(2021)]{2021MNRAS.506.5494P} Poulain, M., Marleau, F.~R., Habas, R., et al.\ 2021, MNRAS, 506, 5494. 


\bibitem[Richardson et al.(2011)]{2011ApJ...732...76R} Richardson, J.~C., Irwin, M.~J., McConnachie, A.~W., et al.\ 2011, APJ, 732, 76. 

\bibitem[Sales et al.(2022)]{2022NatAs...6..897S} Sales, L.~V., Wetzel, A., \& Fattahi, A.\ 2022, Nature Astronomy, 6, 897. 

\bibitem[Sales \& Navarro(2023)]{2023NatAs.tmp...59S} Sales, L.~V. \& Navarro, J.~F.\ 2023, Nature Astronomy. 


\bibitem[Sand et al.(2014)]{2014ApJ...793L...7S} Sand, D.~J., Crnojevi{\'c}, D., Strader, J., et al.\ 2014, APJL, 793, L7. 


\bibitem[Santos-Santos et al.(2020)]{2020MNRAS.499.3755S} Santos-Santos, I.~M., Dom{\'\i}nguez-Tenreiro, R., \& Pawlowski, M.~S.\ 2020, MNRAS, 499, 3755. 

\bibitem[Sawala et al.(2022)]{2022NatAs.tmp..273S} Sawala, T., Cautun, M., Frenk, C., et al.\ 2022, Nature Astronomy. 


\bibitem[Sohn et al.(2020)]{2020ApJ...901...43S} Sohn, S.~T., Patel, E., Fardal, M.~A., et al.\ 2020, APJ, 901, 43. 


\bibitem[Tully et al.(2015)]{2015ApJ...802L..25T} Tully, R.~B., Libeskind, N.~I., Karachentsev, I.~D., et al.\ 2015, APJL, 802, L25. 

\bibitem[Walsh et al.(2007)]{2007ApJ...662L..83W} Walsh, S.~M., Jerjen, H., \& Willman, B.\ 2007, APJL, 662, L83. 


\bibitem[Willman et al.(2005)]{2005ApJ...626L..85W} Willman, B., Dalcanton, J.~J., Martinez-Delgado, D., et al.\ 2005, APJL, 626, L85. 


\bibitem[Xu et al.(2023)]{2023arXiv230300441X} Xu, Y., Kang, X., \& Libeskind, N.~I.\ 2023, arXiv:2303.00441.

\end{thebibliography}





\end{document}